\documentclass[12pt]{article}

\usepackage{amsmath}
\usepackage{graphicx,psfrag,epsf}
\usepackage{enumerate}
\usepackage{booktabs}
\usepackage{natbib}
\usepackage{url} 
\usepackage{amsthm}
\theoremstyle{definition}
\newtheorem{assumption}{Assumption}
\newtheorem{proposition}{Proposition}
\newtheorem{condition}{Condition}
\newtheorem{theorem}{Theorem}
\newtheorem{lemma}{Lemma}
\usepackage{booktabs}
\usepackage{framed}  
\usepackage{caption}
\usepackage{pgfplots}
\usepgfplotslibrary{dateplot}
\usetikzlibrary{snakes}
\usepackage{rotating}
\usepackage{amsfonts}
\usepackage{multirow, booktabs}
\usepackage{cleveref}
\usepackage{subfig}
\usepackage[ruled,vlined]{algorithm2e}
\usepackage[T1]{fontenc}
\usepackage[utf8]{inputenc}
\usepackage{authblk}
\usepackage[multiple]{footmisc}
\usepackage{blindtext,titlefoot}
\usepackage{sectsty}
\usepackage{xcolor}
\usepackage{tikz}
\usetikzlibrary{positioning, shapes.geometric}

\usepackage{tabularx}
\usepackage{ragged2e} 
\newcolumntype{L}{>{\RaggedRight}X} 
\usepackage{lipsum} 

\usepackage[ruled,vlined]{algorithm2e}
\usepackage{amsmath, amssymb}
\usepackage{amsfonts, multirow, epsfig, subfig, floatrow}
\usepackage{graphicx, pdflscape, verbatim, enumerate, colortbl, setspace}
\usepackage{setspace, color,bm}
\usepackage[normalem]{ulem}
\usepackage{cite}
\usepackage{multirow}
\usepackage{booktabs,array}
\usepackage{url}
\usepackage{bbm}

\newtheorem{Remark}{Remark}

\makeatletter
\newcommand*{\rom}[1]{\expandafter\@slowromancap\romannumeral #1@}
\makeatother

\begin{document}

\def\spacingset#1{\renewcommand{\baselinestretch}%
{#1}\small\normalsize} \spacingset{1}

\sectionfont{\bfseries\large\sffamily}%
%
\newcommand*\emptycirc[1][1ex]{\tikz\draw (0,0) circle (#1);} 
\newcommand*\halfcirc[1][1ex]{%
  \begin{tikzpicture}
  \draw[fill] (0,0)-- (90:#1) arc (90:270:#1) -- cycle ;
  \draw (0,0) circle (#1);
  \end{tikzpicture}}
\newcommand*\fullcirc[1][1ex]{\tikz\fill (0,0) circle (#1);} 

\subsectionfont{\bfseries\sffamily\normalsize}%
%


\def\spacingset#1{\renewcommand{\baselinestretch}%
{#1}\small\normalsize} \spacingset{1}

\begin{center}
    \Large \bf A Non-Bipartite Matching Framework for Difference-in-Differences with General Treatment Types
\end{center}



\begin{center}
  \large  $\text{Siyu Heng}^{1,*}$, $\text{Yuan Huang}^{2,*}$, and $\text{Hyunseung Kang}^{3}$
\end{center}

\begin{center}
   \large \textit{$^{1}$Department of Biostatistics, New York University}
\end{center}

\begin{center}
   \large \textit{$^{2}$Department of Applied Statistics, Social Science, and Humanities, New York University}
\end{center}

\begin{center}
   \large \textit{$^{3}$Department of Statistics, University of Wisconsin-Madison }
\end{center}

\bigskip

\begin{abstract}

Difference-in-differences (DID) is one of the most widely used causal inference frameworks in observational studies. However, most existing DID methods are designed for binary treatments and cannot be readily applied to non-binary treatment settings. Although recent work has begun to extend DID to non-binary (e.g., continuous) treatments, these approaches typically require strong additional assumptions, including parametric outcome models or the presence of idealized comparison units with (nearly) static treatment levels over time (commonly called “stayers’’ or “quasi-stayers’’). In this technical note, we introduce a new non-bipartite matching framework for DID that naturally accommodates general treatment types (e.g., binary, ordinal, or continuous). 
Our framework makes three main contributions. First, we develop an optimal non-bipartite matching design for DID that jointly balances baseline covariates across comparable units (reducing bias) and maximizes contrasts in treatment trajectories over time (improving efficiency). Second, we establish a post-matching randomization condition, the design-based counterpart to the traditional parallel-trends assumption, which enables valid design-based inference. Third, we introduce the sample average DID ratio, a finite-population-valid and fully nonparametric causal estimand applicable to arbitrary treatment types. Our design-based approach that preserves the full treatment-dose information, avoids parametric assumptions, does not rely on the existence of stayers or quasi-stayers, and operates entirely within a finite-population framework, without appealing to hypothetical super-populations or outcome distributions. 

\end{abstract}

\noindent%

\let\thefootnote\relax\footnotetext{$^{*}$Siyu Heng and Yuan Huang contributed equally to this work.}

\let\thefootnote\relax\footnotetext{This working manuscript reports preliminary results from our ongoing research on this topic. Additional simulation studies, data applications, methodological developments, and detailed results are still in progress and will be incorporated in future versions. This document is therefore not a final version of our work.}

\spacingset{1.73} 

\section{Introduction}

Difference-in-differences (DID) is one of the most widely used quasi-experimental designs in observational studies. In the binary-treatment settings, DID estimates causal effects by comparing the changes in outcome over time between the treated and control groups. Estimation and inference (i.e., uncertainty quantification) are also well-established under DID with binary treatments \citep{card1993minimum, bertrand2004much, abadie2005semiparametric, athey2006identification, angrist2009mostly, imai2023matching, daw2018matchingregression, ham2024benefits}. However, when the treatment of interest is inherently non-binary (e.g., continuous or ordinal), the canonical binary-treatment DID methods cannot be directly applied. A common workaround in applied research is to dichotomize the originally continuous or ordinal treatment based on a heuristic cut-off (e.g., the median treatment dose) and then apply binary-treatment DID methods. However, recent work has shown that this empirical strategy can lead to severely biased and misleading causal conclusions due to discarding key treatment dose information and omitting underlying dose-response relationships \citep{de2022difference, callaway2024difference, wang2024re}. 

There is fast-growing literature on developing DID methods for non-binary (e.g., continuous) treatments \citep{normington2019bayesian, de2022difference, callaway2024difference, de2024difference, zhang2024continuous, wang2024re, han2019causal, haddad2024difference, hettinger2025multiply}. However, all of these existing methods rely on either some parametric structural models (e.g., linear dose-response models) or a sufficient number of idealized control units with (nearly) static treatment values over time (often referred to as ``stayers'' or ``quasi-stayers''). In many real-world datasets, neither of these two assumptions are plausible: the underlying dose-response relationships are usually non-linear, and the number of ``stayers'' or ``quasi-stayers'' is often scarce when there is an evident, overall trend in the treatment dose. In addition, all existing DID frameworks that can handle non-binary treatments are model-based (i.e., super-population-based), which assume the study units are independently and identically distributed (i.i.d.) realizations from some super-population outcome model. However, in a number of real-world DID studies, the study units are only well-defined as a finite population (e.g., all U.S. states or counties) \citep{abadie2020sampling, zhang2023social}. Recent work has suggested that directly applying super-population (model-based) inference methods in finite-population (design-based) settings can reduce both interpretability (e.g., producing ill-defined causal estimands) and validity (e.g., inflating Type-I error rates) \citep{young2019channeling, abadie2020sampling, athey2022design, rambachan2025design}.

To address these limitations, in this technical note, we propose a new framework for DID with general treatment types (binary, ordinal, or continuous). Building on statistical non-bipartite matching and randomization-based inference, this new DID framework can accommodate valid inference for causal effects without relying on parametric effect models or the existence of ``stayers'' or ``quasi-stayers.'' In addition, our framework is design-based and finite-population-valid and does not invoke hypothetical super-populations or i.i.d. samples from outcome distributions. The rest of this technical note is organized as follows:

\begin{itemize}
    \item In Section~\ref{subsec: matching design}, we describe a DID design based on optimal non-bipartite matching \citep{lu2001matching, lu2011optimal, baiocchi2012near, greevy2023optimal}, which forms matched pairs of study units with the same or similar covariates (for reducing bias) but differing treatment trajectories over time (for improving efficiency). In contrast to the traditional matching design for DID, our framework accommodates non-binary treatment types without losing treatment dose information (e.g., avoiding dose information loss due to dichotomization or coarsening). Also, when there are only two treatment doses (i.e., when the treatment is binary), our non-bipartite matching DID design will reduce to the classic binary-treatment matching (bipartite matching) DID design \citep{heckman1997matching, smith2005does, daw2018matching, daw2018matchingregression, heng2021relationship, imai2023matching, ham2024benefits}.

    \item In Section~\ref{subsec: DID condition}, based on the non-bipartite matching design and the assumption of no unmeasured time-varying confounding, we derive the key DID condition considered in our framework, namely the \textit{post-matching} randomization condition. Unlike the conventional parallel trend assumption and its extensions or variants \citep{card1993minimum, bertrand2004much, abadie2005semiparametric, athey2006identification, angrist2009mostly, wing2018designing, goodman2021difference, callaway2021difference, imai2023matching, daw2018matchingregression, callaway2024difference}, the derived post-matching randomization condition does not rely on any super-population outcome distributions or models, which paves the way for developing the downstream design-based (randomization-based) inference methods under our DID framework. Moreover, to our knowledge, existing design-based DID formulations \citep{athey2022design, rambachan2025design} focus on binary treatment indicators, whereas our post-matching randomization condition is defined for general treatment types, including binary, multi-valued, and continuous doses. In short, the post-matching randomization condition can be viewed as a design-based counterpart to the parallel trends condition to settings with any treatment types.
    

    \item In Section~\ref{subsec: estimands}, building on the proposed new DID design, we introduce a new causal estimand, called the \textit{sample average DID ratio}, to provide a meaningful and interpretable summary of the overall causal effects in DID with general treatment types. Unlike the existing causal estimands in DID \citep{card1993minimum, bertrand2004much, abadie2005semiparametric, athey2006identification, angrist2009mostly, athey2022design, rambachan2025design, callaway2024difference, de2022difference, de2024difference, wang2024re}, which either require invoking hypothetical super-populations (or outcome distributions) or are restricted to binary treatments, the proposed causal estimand (i.e., the sample average DID ratio) is finite-population-valid, non-parametric, and universally applicable to arbitrary treatment types (e.g., binary, ordinal, or continuous). 

    \item In Section~\ref{subsec: inference}, building on the derived key DID condition (i.e., the post-matching randomization condition), we propose a design-based (randomization-based), non-parametric inference framework for meaningful causal estimands (e.g., the sample average DID ratio). Unlike existing design-based inference frameworks in DID \citep{athey2022design, rambachan2025design}, which are restricted to binary treatments, our framework is universally applicable to binary and non-binary (e.g., ordinal or continuous) treatments. 

\end{itemize}

\section{A Non-bipartite Matching Design for DID with General Treatment Types}

\subsection{The Proposed Matching Design }\label{subsec: matching design}


We develop a new DID study design for \emph{general} (binary, ordinal, continuous) treatments using the techniques of \emph{optimal non-bipartite matching} \citep{lu2001matching, baiocchi2010building, lu2011optimal, greevy2023optimal}. The design encodes two targets: (i) balance baseline covariates between the comparison units to control for bias due to measured confounding bias, and (ii) maximize the contrast in change in treatment dose over time between the comparison units to improve statistical power for detecting the potential causal effect. Unlike classical bipartite matching for binary-treatment DID \citep{heckman1997matching, smith2005does, daw2018matching, daw2018matchingregression, heng2021relationship, imai2023matching, ham2024benefits}, our non-bipartite matching DID design does not require dichotomizing or coarsening an originally non-binary treatment to form matched units (i.e., comparison units), and will reduce to the classical bipartite matching when the treatment is binary. This flexibility allows us to fully preserve treatment dose information, which (i) allows the causal effect estimands and corresponding statistical inference to be dose-responsive (i.e., allows the causal effect to change with magnitudes of treatment dose) and (ii) avoids suboptimal post-matching covariate balance due to dichotomization.

For clarity, we consider two periods $t\in\{0,1\}$, where $t=0$ denotes the early period and $t=1$ denotes the late period. Let $N$ units be indexed by $n=1,\dots,N$. For unit $n$, let $Z_{n}^{0}$ and $Z_{n}^{1}$ denote the treatment value (e.g., continuous, ordinal, or binary) at $t=0$ and $t=1$, respectively. Then, the treatment change over time is denoted as $\Delta^{Z}_n = Z_{n}^{1}-Z_{n}^{0}$. Let $\mathbf{x}_n$ and $u_{n}$ denote the collection of baseline covariates and the unobserved covariate, respectively. Here, we allow observed covariates $\mathbf{x}_{n}$ to include either time-invariant or time-varying covariates, or both, and we omit the time index $t$ for simplicity of notations. The goal is to partition the study units into $I$ disjoint pairs ($N=2I$ if $N$ is even) such that pairs are \emph{well-balanced} in observed covariates $\mathbf{x}$ and \emph{well-separated} in $\Delta^{Z}$.

Specifically, the proposed DID design can be formulated as the following non-bipartite matching procedure: consider a graph consisting of a pair of sets $([N], E)$, where $[N] = \{1,\dots, N\}$ represents the set of indexes of $N$ study units (each index/unit can be considered as a node in the graph), and $E = \{(n_{1}, n_{2}): n_{1}<n_{2} \}$ represents the set of undirected edges that connect pairs of units (nodes). Under the optimal non-bipartite matching framework, there is no predefined treatment or control group. Therefore, the only constraints for edges in $E$ are: 1) $(n, n)\notin E$ for all $n=1,\dots, N$ (i.e., a unit can not be matched with itself); and 2) if $(n_{1}, n_{2})\in E$ for some $n_{1}$ and $n_{2}$, then for any $n_{3} \neq n_{2}$, we have $(n_{1}, n_{3})\notin E$ (i.e., matching without replacement). Let $\mathcal{E}$ denote the collection of all possible sets of edges $E$ that satisfy these two constraints. The optimal set of matched pairs $E^{*} \in \mathcal{E}$ minimizes the total distances (based on some prespecified distance metric $d$) between matched individuals. Formally, the corresponding optimal non-bipartite matching problem tries to solve
\begin{equation}\label{eqn: nbp matching}
    E^{*} = \underset{E \in \mathcal{E}}{\text{argmin}}\sum_{(n_{1}, n_{2})\in E} d(n_{1}, n_{2}),
\end{equation}
where $d(n_{1}, n_{2})$ is some prespecified distance metric between units (nodes) $n_{1}$ and $n_{2}$. There are different ways to define the distance metric to incorporate the two major design considerations: bias (pairing units with similar $\mathbf{x}$) and efficiency (pairing units with evidently different $\Delta^{Z}$). In this technical note, we focus on two sensible and easy-to-implement choices that can encode these two considerations:

\begin{itemize}
    \item \textbf{Ratio-type distance:} For any edge $(n_1,n_2)$, we define
\[d_{\mathrm{ratio}}(n_1,n_2)
\;=\;\frac{\delta_x\!\big(\mathbf x_{n_1},\mathbf x_{n_2}\big)}{\delta_z\!\big(\Delta^{Z}_{n_1},\Delta^{Z}_{n_2}\big)+\epsilon}\,,\]
where $\delta_x$ is a prespecified covariate distance and $\delta_z$ measures treatment-change separation. In applications, we could take $\delta_x$ and $\delta_{z}$ to be the Mahalanobis distance or its rank-based version, though other robust choices are admissible \citep{rosenbaum2020design,rosenbaum2020modern}. This ratio-type distance favors edges that are simultaneously well balanced in $\mathbf x$ and far apart in $\Delta^{Z}$. In practice, we recommend adding a small regularization constant $\epsilon>0$ into $d_{\mathrm{ratio}}(n_1, n_2)$, which prevents the ratio-type distance from blowing up among pairs with extremely close treatment-change $\Delta^{Z}$.  

\item \textbf{Penalty-based distance:} To ensure sufficient covariate balance and contrast in treatment change $\Delta^{Z}$, augment the ratio with hinge penalties:
\[
d_{\mathrm{pen}}(n_{1}, n_{2})=\delta_{x}\!\big(\mathbf x_{n_1},\mathbf x_{n_2}\big)+M\cdot \mathbf{1}\{|\Delta^{Z}_{1} -\Delta^{Z}_{2}|\leq \xi\},
\]
where $M$ is a prespecified large number and $\xi>0$ is some prespecified threshold of the tolerable difference in $\Delta^{Z}$. This strategy was first adopted in a recent applied research \citep{wang2024re}. 
\end{itemize}

We will apply Derigs' algorithm \citep{derigs1988solving} to solve the optimal non-bipartite matching problem (\ref{eqn: nbp matching}), which partitions the $N$ units into $I$ independent matched pairs ($N=2I$ if $N$ is even), minimizing the total within-pair distance $\sum d(n_{1},n_{2})$ across all possible pairings. The proposed design turns DID with general treatments into a design-based matching problem that explicitly \emph{balances} baseline covariates and \emph{separates} treatment changes. It preserves the full richness of treatment dose information, avoids dichotomization or coarsening, and prepares the ground for the design-based identification and inference introduced in later sections. When the treatment $Z$ is binary (i.e., when there are two treatment doses), the proposed non-bipartite matching DID design will reduce to the classical bipartite matching design for DID. Also, the proposed non-bipartite matching DID design can be naturally extended to multiple time periods by generalizing non-bipartite matching, e.g., by replacing $\delta_{z}(\Delta_{n_1}^{Z},\Delta_{n_2}^{Z})$ with a trajectory distance $\widetilde{\delta}_{z}(\mathbf{Z}_{n_1},\mathbf{Z}_{n_2})$). 


\subsection{Formulation of the Key DID Condition}\label{subsec: DID condition}

After applying the non-bipartite matching DID design, we form units into $I$ independent matched pairs. For unit $j$ in matched pair $i$ ($j=1,2$ and $i=1,\dots, I$), we let $Z_{ij}^{t=0}$ (or $Z_{ij}^{t=1}$) denote the treatment dose at $t=0$ (or $t=1$), $\mathbf{Z}_{ij}=(Z_{ij}^{t=0}, Z_{ij}^{t=1})$ the treatment dose trajectory, $\Delta_{ij}^{Z}=Z_{ij}^{t=1}-Z_{ij}^{t=0}$ the change in treatment dose over time, $\mathbf{x}_{ij}$ the baseline covariates (measured confounders), $u_{ij}$ an unmeasured confounder, $Y_{ij}^{t=0}$ (or $Y_{ij}^{t=1}$) the observed outcome at $t=0$ (or $t=1$), $\mathbf{Y}_{ij}=(Y_{ij}^{t=0}, Y_{ij}^{t=1})$ the observed outcome trajectory, and $\Delta_{ij}^{Y}=Y_{ij}^{t=1}-Y_{ij}^{t=0}$ the change in outcome over time. Within each pair, matching makes baseline covariates equal or close (i.e., $\mathbf x_{i1}= \mathbf x_{i2}$ or $\mathbf x_{i1}\approx \mathbf x_{i2}$). For each matched pair $i$, let $\Delta^Z_{i,**}=\max\{\Delta^Z_{i1},\Delta^Z_{i2}\}$ (or $\Delta^Z_{i,*}=\min\{\Delta^Z_{i1},\Delta^Z_{i2}\}$) denote the larger (or the smaller) of the two observed changes in treatment dose from the two units $i1$ and $i2$, and let $\mathbf z_{i,**}=(z^{t=0}_{i,**},z^{t=1}_{i,**})$ (or $\mathbf z_{i,*}=(z^{t=0}_{i,*},z^{t=1}_{i,*})$)
be the treatment dose trajectory of the unit whose treatment dose change equals $\Delta^Z_{i,**}$ (or $\Delta^Z_{i,*}$).
Thus, conditional on matching, the potential within-pair treatment trajectories $(\mathbf Z_{i1},\mathbf Z_{i2})$ equals either $(\mathbf z_{i,**},\mathbf z_{i,*})$ or $(\mathbf z_{i,*},\mathbf z_{i,**})$. Let $\mathcal Z = \big\{ \mathbf{Z}=(\mathbf{Z}_{11},\dots, \mathbf{Z}_{I2}): (\mathbf Z_{i1},\mathbf Z_{i2})=(\mathbf z_{i,**}, \mathbf z_{i,*}) \text{ or } (\mathbf z_{i,*},\mathbf z_{i,**})\ \text{for all }i\big\}$ denote the set of all \emph{post-matching} assignments of treatment dose trajectories, where $|\mathcal Z|=2^I$.

Under the potential outcomes framework \citep{neyman1923application, rubin1974estimating}, we define the pair-specific potential changes in outcome corresponding to the larger and smaller treatment change: $\Delta^{Y}_{ij,**}=Y_{ij}^{t=1}(z^{t=1}_{i,**})-Y_{ij}^{t=0}(z^{t=0}_{i,**})$, $\Delta^{Y}_{ij,*}=Y_{ij}^{t=1}(z^{t=1}_{i,*})-Y_{ij}^{t=0}(z^{t=0}_{i,*})$. Let $\mathbf X=(\mathbf x_{11},\dots,\mathbf x_{I2})$ and $\mathcal F = \big\{ \mathbf X,\
\{\Delta^{Y}_{ij,**},\Delta^{Y}_{ij,*}\}_{i=1,\dots,I;\, j=1,2} \big\}$. We adopt the following design-based assumptions:
\begin{assumption}[Consistency, SUTVA, and no anticipatory effects]\label{assump: consistency}
    We have $Y_{ij}^t=Y_{ij}^t(Z_{ij}^t)$ for all $i=1,\dots, I$, $j=1,2$, and $t=0, 1$.
\end{assumption}
\begin{assumption}[No unmeasured time-varying confounding]\label{assump: ignorability}
    The additive unmeasured confounding $u_{ij}$, if it exists, is time-invariant. 
\end{assumption}

\begin{Remark}
    Assumption~\ref{assump: ignorability} rules out unmeasured confounders whose effects vary over time, so any additive unmeasured term $u_{ij}$, if present, is constant across periods for a given unit. Consequently, although different units may have different levels of unobserved confounding variables, these differences affect only the level of the outcome and not its evolution over time. As a result, unobserved heterogeneity cannot generate differential changes in outcomes across units.
    
\end{Remark}


\begin{assumption}[Positivity/overlap within pairs.]\label{assump: positivity}
 For each unit $ij$, we have $P(\mathbf{Z}_{ij}=\mathbf{z}_{i,**}\mid \mathcal{F})>0$ and $P(\mathbf{Z}_{ij}=\mathbf{z}_{i,**}\mid \mathcal{F})>0$ (here $P$ denotes conditional probability for discrete treatments and denotes conditional density for continuous treatments). 
\end{assumption}

\begin{proposition}[Post-matching randomization under Assumptions~\ref{assump: consistency}--\ref{assump: positivity}]
Under Assumptions~\ref{assump: consistency}--\ref{assump: positivity} and conditional on matching on observed covariates ($\mathbf{x}_{i1}=\mathbf{x}_{i2}$ or $\mathbf{x}_{i1}\approx \mathbf{x}_{i2}$), we have: for each matched pair $i = 1,\dots,I$,
\begin{equation}\label{eq:randomization_condition}
P(\mathbf{Z}_{i1} = \mathbf{z}_{i,**},\, \mathbf{Z}_{i2} = \mathbf{z}_{i,*} \mid \mathcal{F}, \mathcal{Z})
= P(\mathbf{Z}_{i1} = \mathbf{z}_{i,*},\, \mathbf{Z}_{i2} = \mathbf{z}_{i,**} \mid \mathcal{F}, \mathcal{Z})
= \frac{1}{2}.
\end{equation}
We refer to \eqref{eq:randomization_condition} as the \emph{post-matching randomization condition}.
\end{proposition}

The post-matching randomization condition (\ref{eq:randomization_condition}) is the cornerstone of design-based (i.e., randomization-based) inference in our DID framework. It arises because, under the assumption of time-invariant unmeasured confounding (Assumption~\ref{assump: ignorability}), the potential outcome changes $(\Delta ^{Y}_{ij,**}, \Delta^{Y}_{ij,*})$ depend only on treatment trajectories and observed baseline covariates $\mathbf{X}$, not on the unobserved confounder $u_{ij}$ (since $u_{ij}$ cancels in the outcome difference $\Delta^{Y}_{ij}$). In other words, there is no information about $u_{ij}$ in the potential outcome changes $(\Delta ^{Y}_{ij,**}, \Delta^{Y}_{ij,*})$. Therefore, conditional on matched observed covariates and potential outcome changes $(\Delta ^{Y}_{ij,**}, \Delta^{Y}_{ij,*})$ (i.e., the information in $\mathcal{F}$), as well as the set of feasible treatment trajectories within each pair (i.e, the information in $\mathcal{Z}$), the assignment of which unit receives trajectory $\mathbf{z}_{i,**}$ versus $\mathbf{z}_{i,*}$ is label-exchangeable, yielding equal probability to both possible orderings. This condition is a design-based counterpart to the various versions of parallel trends assumptions \citep{card1993minimum, bertrand2004much, abadie2005semiparametric, athey2006identification, angrist2009mostly, wing2018designing, goodman2021difference, callaway2021difference, imai2023matching, daw2018matchingregression, callaway2024difference} in traditional super-population DID and provides the basis for constructing finite-population-valid, randomization-based inference without relying on hypothetical super-populations or assumptions on outcome distributions. In addition, enabled by the proposed non-bipartite matching DID design, condition (\ref{eq:randomization_condition}) extends recent work on design-based (randomization-based) DID conditions \citep{athey2022design, rambachan2025design} from binary treatments to general treatments.


\section{Design-based DID Analysis under General Treatment Types}

In this section, based on the non-bipartite matching DID design, we propose a design-based inference framework. In contrast to the well-established super-population DID inference frameworks, either for binary or non-binary treatments (e.g., \citealp{card1993minimum, abadie2005semiparametric, athey2006identification, angrist2009mostly, goodman2021difference, callaway2021difference, imai2023matching, callaway2024difference}), our estimands are well defined for a finite population, and our inference methods do not rely on hypothetical super-populations or assumptions on outcome distribution or modeling, and do not require the existence of ``stayers'' or ``quasi-stayers.'' In addition, unlike existing design-based DID inference frameworks \citep{athey2022design, rambachan2025design}, which are restricted to binary treatments, our framework supports general treatment types (e.g., binary, ordinal, or continuous).

\subsection{Finite-Population Causal Estimands}\label{subsec: estimands}

The first step of building our design-based inference framework is to introduce finite-population-valid and scientifically meaningful causal estimands. For unit $j$ in matched pair $i$ (where $j \in \{1, 2\}$ and $i = 1, \ldots, I$), we introduce the unit-specific DID ratio $\tau_{ij}$, defined as $\tau_{ij} = \frac{\Delta^{Y}_{ij,**} - \Delta^{Y}_{ij,*}}{\Delta^{Z}_{i,**} - \Delta^{Z}_{i,*}}$, where $\Delta^{Y}_{ij,**}$ and $\Delta^{Y}_{ij,*}$ denote the potential outcome changes under the larger and smaller change in treatment doses, respectively (recall that $\Delta^{Z}_{i,**} - \Delta^{Z}_{i,*} > 0$). That is, $\tau_{ij}$ measures the rate of change in outcomes with respect to changes in treatment dose, and a positive (or negative) value of $\tau_{ij}$ indicates that a larger increase in treatment dose corresponds to a larger (or smaller) increase in outcome, suggesting a positive (or negative) causal effect. Aggregating over all matched pairs and units, we define the \textit{sample average DID ratio}:
\begin{equation}\label{eq:sample_avg_did_ratio}
\tau = \frac{1}{2I} \sum_{i=1}^{I} \sum_{j=1}^{2} \tau_{ij} = \frac{1}{2I} \sum_{i=1}^{I} \sum_{j=1}^{2} \frac{\Delta^{Y}_{ij,**} - \Delta^{Y}_{ij,*}}{\Delta^{Z}_{i,**} - \Delta^{Z}_{i,*}}.
\end{equation}
We give some remarks on the estimand $\tau$ (i.e., the sample average DID ratio) defined in (\ref{eq:sample_avg_did_ratio}). First, $\tau$ is finite-population-exact and nonparametric, so it does not rely on any outcome distributional or effect modeling assumptions. Second, $\tau$ works seamlessly and universally across different treatment types (binary, ordinal, or continuous). When the treatment is binary (i.e., there are only two treatment doses) with $(\Delta^{Z}_{i,**}, \Delta^{Z}_{i,*}) = (1, 0)$, $\tau$ reduces to the classical sample average treatment effect in finite-population DID settings \citep{athey2022design, rambachan2025design}. To our knowledge, the sample average DID ratio $\tau$ is the first finite-population-valid, nonparametric causal estimand in DID beyond the binary treatment paradigm. Third, $\tau$ does not require the existence of ``stayers'' or ``quasi-stayers'' (i.e., idealized control units with (nearly) static treatment dose over time). That is, neither $\Delta^{Y}_{ij,**}$ nor $\Delta^{Y}_{ij,*}$ needs to equal or be close to zero, which greatly extends the applicability of DID studies in continuous treatment settings, given that the existing nonparametric DID frameworks under continuous treatments rely on the sufficient number of  ``stayers'' or ``quasi-stayers'' \citep{de2022difference, callaway2024difference, de2024difference}. To build further intuition, consider a hypothetical special case with a linear effect of the treatment dose on the outcome. Suppose there exists
a scalar $\beta$ such that, for every unit $ij$, we have $\Delta^Y_{ij,**} - \Delta^Y_{ij,*} = \beta \,\bigl(\Delta^Z_{i,**} - \Delta^Z_{i,*}\bigr)$. In this linear effect scenario, each unit-specific DID ratio satisfies $\tau_{ij} = \beta$ and hence the sample average DID ratio reduces to $\tau = \frac{1}{2I}\sum_{i=1}^I \sum_{j=1}^2 \tau_{ij} = \beta$. More generally, if we allow unit-specific slopes $\beta_{ij}$ such that $\Delta^Y_{ij,**} - \Delta^Y_{ij,*} = \beta_{ij} \,\bigl(\Delta^Z_{i,**} - \Delta^Z_{i,*}\bigr)$, then $\tau$ is simply the sample average of these unit-level effects,
$\tau = \frac{1}{2I}\sum_{i=1}^I \sum_{j=1}^2 \beta_{ij}$. Our framework and estimand do not
require any such linear or parametric effect model; this example is purely for interpretation, and it highlights that $\tau$ accommodates arbitrary unit-level heterogeneity in the dose-response relationship.

The sample average DID ratio estimand extends naturally to a broad family of nonparametric estimands of the form:
\begin{equation}
\label{eq:general_estimand}
\theta = \frac{1}{2I} \sum_{i=1}^{I} \sum_{j=1}^{2} \frac{\psi_y(\Delta^{Y}_{ij,**}) - \psi_y(\Delta^{Y}_{ij,*})}{\psi_z(\Delta^{Z}_{i,**}, \Delta^{Z}_{i,*})},
\end{equation}
where $\psi_y$ and $\psi_z$ are arbitrary, user-defined functions. For instance, setting $\psi_y(y) = \log(y)$ and/or $\psi_z(z_1, z_2) = \log(z_1 / z_2)$ yields a DID-ratio-type estimand defined at the logarithmic scale of outcome and/or treatment dose. This flexibility allows practitioners to target estimands tailored to their scientific questions while maintaining the finite-population, design-based inference framework.

\subsection{Design-based inference}\label{subsec: inference}

In Section~\ref{subsec: inference}, we present a design-based framework for estimation and inference of the sample average DID ratio $\tau$ defined in (\ref{eq:sample_avg_did_ratio}). The framework can be readily generalized to infer $\theta$ defined in the general estimand defined in (\ref{eq:general_estimand}) by replacing $\Delta^{Y}_{ij}$ with $\psi_y(\Delta^{Y}_{ij})$ and replacing $\Delta^{Z}_{i,**} - \Delta^{Z}_{i,*}$ with $\psi_z(\Delta^{Z}_{i,**}, \Delta^{Z}_{i,*})$. 

For the sample average DID ratio $\tau$, we defined the following \textit{DID ratio estimator}:  
\begin{equation}
\label{eq:did_ratio_estimator}
\widehat{\tau} = \frac{1}{I} \sum_{i=1}^{I} \widehat{\tau}_i, \quad \text{where} \quad \widehat{\tau}_i = \frac{\Delta^{Y}_{i1} - \Delta^{Y}_{i2}}{\Delta^{Z}_{i1} - \Delta^{Z}_{i2}}.
\end{equation}
Here, recall that $\Delta^Y_{ij} = Y_{ij}^{t=1} - Y_{ij}^{t=0}$ denotes the observed change in outcome for unit $j$ in pair $i$ between $t=0$ and $t=1$, so that $\Delta^Y_{i1} - \Delta^Y_{i2}
  = (Y_{i1}^{t=1} - Y_{i1}^{t=0}) - (Y_{i2}^{t=1} - Y_{i2}^{t=0})$ is the DID in outcomes within pair $i$. Also, recall that $\Delta^Z_{ij} = Z_{ij}^{t=1} - Z_{ij}^{t=0}$ denotes the observed change in treatment dose for unit $j$ in pair $i$, so $\Delta^Z_{i1} - \Delta^Z_{i2}=(Z_{i1}^{t=1} - Z_{i1}^{t=0}) - (Z_{i2}^{t=1} - Z_{i2}^{t=0})$ is the DID in treatment doses within pair $i$. Thus, $\widehat{\tau}_i$ measures the within-pair ratio of DID in outcomes to DID in treatment doses, and $\widehat{\tau}$ aggregates this DID ratio across all $I$ matched pairs. Under the post-matching randomization condition \eqref{eq:randomization_condition}, we can prove that $\widehat{\tau}$ is an unbiased estimator for $\tau$.
\begin{proposition}
\label{prop:unbiasedness}
Under the post-matching randomization condition \eqref{eq:randomization_condition}, we have $E(\widehat{\tau} \mid \mathcal{F}, \mathcal{Z}) = \tau$.
\end{proposition}

We present a simulation study comparing three DID estimators in the continuous treatment case: the dichotomized DID estimator (obtained by dichotomizing a continuous treatment at the median), the parametric DID estimator (based on fitting a linear model to outcome changes), and our proposed nonparametric DID ratio estimator $\widehat{\tau}$ (based on the proposed non-bipartite matching DID design). We generated $N = 2{,}000$ independent units with covariates $\mathbf{x} = (x_1, x_2, x_3)\sim N(\mathbf{0}, \mathbf{I}_{3\times 3})$ and an unobserved confounder $u \sim N(0,1)$. Treatment dose trajectories $(Z^{t=0}, Z^{t=1})$ were constructed as $Z^t = f_z^t(\mathbf{x}) + 0.3u + \epsilon_z^t$ with $\epsilon_z^t \sim N(0,1)$, and outcome trajectories $(Y^{t=0}, Y^{t=1})$ as $Y^t = \beta Z^t + f_y^t(\mathbf{x}) + 0.2u + \epsilon_y^t$ with $\epsilon_y^t \sim N(0,1)$. The functions $f_z^t(\mathbf{x})$ and $f_y^t(\mathbf{x})$ included linear terms in $(x_1, x_2, x_3)$ and nonlinear terms: $x_1 x_3$, $x_2 x_3$, $\sin(x_1)$, $\cos(x_2)$, and $\exp(x_3/3)$.

\begin{table}[htbp]
\centering
\footnotesize
\caption{Simulated estimation biases under different effect sizes.}
\label{tbl:did_comparison}
\begin{tabular}{lc|lc|lc|lc}
\hline
\multicolumn{2}{c|}{\textbf{$\beta = 1.5$}} & \multicolumn{2}{c|}{\textbf{$\beta = 2$}} & \multicolumn{2}{c|}{\textbf{$\beta = 2.5$}} & \multicolumn{2}{c}{\textbf{$\beta = 3$}} \\
\textbf{Estimator} & \textbf{Bias} & \textbf{Estimator} & \textbf{Bias} & \textbf{Estimator} & \textbf{Bias} & \textbf{Estimator} & \textbf{Bias}\\
\hline
Dichotomized DID & 4.27 & Dichotomized DID & 5.35 & Dichotomized DID & 6.42 & Dichotomized DID & 7.49 \\
Parametric DID & 0.38 & Parametric DID & 0.38 & Parametric DID & 0.38 & Parametric DID & 0.37 \\
DID Ratio $\widehat{\tau}$ & 0.06 & DID Ratio $\widehat{\tau}$ & 0.06 & DID Ratio $\widehat{\tau}$ & 0.06 & DID Ratio $\widehat{\tau}$ & 0.05 \\
\hline
\end{tabular}
\end{table}

Table~\ref{tbl:did_comparison} presents simulation results under various effect sizes $\beta$. Our DID ratio estimator $\widehat{\tau}$ substantially reduces bias compared to both the dichotomized and parametric alternatives, even when causal effects are approximately linear (but covariate-response relationships remain nonlinear). Note that when the causal effects are non-linear, the dichotomized DID estimator and the parametric DID estimator do not even have a well-defined causal estimand. In contrast, the proposed DID ratio estimator $\widehat{\tau}$ is still finite-population-valid under arbitrary heterogeneous or non-linear effects. 

Then, we derive a design-based variance estimator for $\widehat{\tau}$ by extending the idea of covariate-adjusted variance estimation from the classical experimental or matching settings \citep{fogarty2018mitigating, zhu2023randomization, frazier2024bias} to the DID settings. Consider any fixed $I \times L$ matrix $Q$ with $L < I$, $Q$. For example, a simplest choice is to set $Q_{I\times 1}=(1,\dots,1)^{T}$. When there is fruitful covariate information, a more efficient choice is to incorporate covariate information and set $Q = (\mathbf{1}_{I \times 1}, \overline{\mathbf{x}}_{1}, \cdots, \overline{\mathbf{x}}_{K})$, where $\overline{\mathbf{x}}_{k} = (\frac{x_{11k} + x_{12k}}{2},\dots, \frac{x_{I1k} + x_{I2k}}{2})$ represents the within-pair average of the $k$-th baseline covariate ($k=1,\dots, K$ and $K<L-1$). The pair-level adjustment $Q$ thus summarizes pre-treatment differences across matched units. Let $H_Q = Q(Q^{T} Q)^{-1}Q^{T}$ be the associated projection matrix with diagonal entries $h_{ii}$ ($i=1,\dots, I$). Define $y_i = \widehat{\tau}_i / \sqrt{1-h_{ii}}$ and $v = (y_1,\dots,y_I)^{T}$, and $\mathcal{I}$ is $I\times I$ identity matrix. The general form of the proposed variance estimator is
\begin{equation}
S^2(Q) \;=\; I^{-2}\, v^{T} (\mathcal{I} - H_Q)\, v.
\end{equation}

To ensure valid design-based inference, we consider the following mild regularity conditions. These conditions can be viewed as generalizations of some commonly used regularity conditions in the matching literature to the DID settings. 

\begin{condition}[No Extreme Pairs]\label{condition: no extreme pairs}

For each matched pair $i$, let $\widehat{\tau}_i^{(1)}
= \frac{\Delta^{Y}_{i1,**} - \Delta^{Y}_{i2,*}}{\Delta^{Z}_{i,**} - \Delta^{Z}_{i,*}}$ and $\widehat{\tau}_i^{(2)}
= \frac{\Delta^{Y}_{i2,**} - \Delta^{Y}_{i1,*}}{\Delta^{Z}_{i,**} - \Delta^{Z}_{i,*}}$ denote the two potential values of the within-pair DID ratio estimator $\widehat{\tau}_{i}$. Define $V_i^{+} = \max\{\widehat{\tau}_i^{(1)},\widehat{\tau}_i^{(2)}\}$, $V_i^{-} = \min\{\widehat{\tau}_i^{(1)},\widehat{\tau}_i^{(2)}\}$, and $M_i = V_i^{+} - V_i^{-}$. As $I\to\infty$, we have $\max_{1 \le i \le I} \frac{M_{i}^2}{\sum_{i=1}^{I} M_{i}^2} \to 0$.

\end{condition}

This type of condition, standard in the matched observational studies literature \citep{zhang2024bridging, zhu2023randomization, frazier2024bias}, ensures that no single matched pair contributes disproportionately to the total variance, so the total contribution from the entire dataset dominates that of any individual pair. A sufficient condition to satisfy Condition~\ref{condition: no extreme pairs} is that $M_i$ is bounded for each $i$. 

\begin{condition}[Bounded Fourth Moments]\label{condition: bounded fourth moments}
    There exists a constant $C_{1} < \infty$ such that for all $I$, we have: $I^{-1}\sum_{i=1}^{I} M_i^4 \le C_{1}$,
$I^{-1}\sum_{i=1}^{I} (\widehat{\tau}_i^{(1)})^4 \le C_{1}$, and $I^{-1}\sum_{i=1}^{I}  (\widehat{\tau}_i^{(2)})^4 \le C_{1}$. Additionally, there exists another constant $C_2 < \infty$ such that $|q_{il}| \leq C_2$ for all $i = 1,..., I, l = 1, ..., L$, where $q_{il}$ denotes the $(i,l)$-th entry of the $I \times L$ matrix $Q$ involved in the variance estimator $S^2(Q)$ .
\end{condition}

\begin{condition}[Convergence of Finite-Population Means]\label{condition: convergence}
    Let $\mu_i = \mathbb{E}\!\left(\widehat{\tau}_i\,\middle|\,\mathcal F,\mathcal Z\right)=\frac{1}{2}(\widehat{\tau}_i^{(1)} + \widehat{\tau}_i^{(2)})$ and $\nu_i^2 = \mathrm{Var}(\widehat{\tau}_i\mid\mathcal{F},\mathcal{Z}) = \frac{(\widehat{\tau}_i^{(1)} - \widehat{\tau}_i^{(2)})^2}{4}$. As $I \to \infty$, we have: (i) $I^{-1}\sum_{i=1}^{I}\mu_i^2$ converges to a finite positive value; (ii) $I^{-1}\sum_{i=1}^{I} \nu_i^2$ converges to a finite positive value; (iii) For all $l=1,...L$, $I^{-1}\sum_{i=1}^{I} \mu_iq_{il}$ converges to a finite positive value; (iv) $I^{-1}Q^{T}Q$ converges to a finite, invertible $L \times L$ matrix $\widetilde{Q}$.
\end{condition}

Then, the following result establishes the validity of the variance estimator $S^{2}(Q)$. 

\begin{lemma}
Under Conditions \ref{condition: bounded fourth moments} and \ref{condition: convergence}, as $I\to\infty$, we have $$\frac{\text{Var}(\widehat{\tau}|\mathcal{F},\mathcal{Z})}{S^{2}(Q)}\xrightarrow{p} 1 - \frac{\lim_{I\to\infty} I^{-1}\boldsymbol{\mu}(\mathcal{I}-H_{Q})\boldsymbol{\mu}^{T}}{\lim_{I\to\infty} I^{-1}\boldsymbol{\mu}(\mathcal{I}-H_{Q})\boldsymbol{\mu}^{T} + \lim_{I\to\infty} I^{-1}\sum_{i=1}^{I}\nu_{i}^{2}}\in (0,1],$$ 
where $\boldsymbol{\mu} = (\mu_{1},\ldots,\mu_{I})$.
\end{lemma}

Invoking the finite-population central limit theorem \citep{li2017general}, we can establish the asymptotic validity of design-based inference based on $\widehat{\tau}$ and $S^{2}(Q)$. 

\begin{theorem}
Under Assumptions~\ref{assump: consistency}--\ref{assump: positivity}, as well as the regularity conditions listed in Conditions~\ref{condition: no extreme pairs}--\ref{condition: convergence}, the coverage rate of confidence interval $[\widehat{\tau} - \Phi^{-1}(1-\alpha/2)S({Q}), \widehat{\tau} + \Phi^{-1}(1-\alpha/2)S({Q})]$ is asymptotically no less than $100(1-\alpha)\%$, where $\Phi$ is the cumulative distribution function of the standard normal distribution and $\alpha \in (0, 0.5)$ is some prespecified significance level.
\end{theorem}

\section{Discussion}

This technical note proposes a new design-based framework for DID with general treatment types, addressing key limitations of existing DID approaches that rely on parametric models or the presence of ``stayers'' or ``quasi-stayers'' with static treatment trajectories to accommodate non-binary treatments. Building on optimal non-bipartite matching, our design preserves full treatment-dose information, balances baseline covariates, and creates meaningful contrasts in treatment changes, all without dichotomization or modeling assumptions. Based on this non-bipartite matching DID design, we derive design-based DID conditions that pave the way for downstream design-based DID analyses, as well as introduce a new family of nonparametric, finite-population-valid causal estimands in DID settings (i.e., the sample average DID ratio and its extensions). Finally, we develop a nonparametric, design-based inference framework for the proposed causal estimands, which does not rely on hypothetical super-populations or the existence of ``stayers'' or ``quasi-stayers.'' To our knowledge, this is the first design-based DID framework capable of accommodating general (binary, ordinal, or continuous) treatment types.

\bibliographystyle{apalike}
\bibliography{references}

\end{document}